\documentclass[a4paper,12pt]{llncs}
\usepackage{amsmath}
\usepackage{amssymb}

\def\x{\boldsymbol{x}}
\def\y{\boldsymbol{y}}
\def\z{\boldsymbol{z}}
\def\g{\boldsymbol{g}}

\title{\bf A correlation-based distance}
\author{Jean-Luc Falcone\inst{1} \and Paul Albuquerque\inst{1,2}}
\institute{
Computer Science Department,
University of Geneva,
1211 Geneva 4, Switzerland
\and 
${\bf I^3}$,
Ecole d'Ing\'enieurs de Gen\`eve, HES-SO,
1202 Geneva, Switzerland \\
e-mail: jean-luc falcone@cui.unige.ch, albuquer@eig.unige.ch
}
\date{\today}

\begin{document}

\maketitle
\begin{abstract}
In this short technical report, we define on the sample space
$\mathbb{R}^D$ a distance between data points which depends on their
correlation. We also derive an expression for the center of mass of a
set of points with respect to this distance.
\end{abstract}

\section{Preliminaries}
For a sample point $\x=(x_1,\dots,x_D)\in\mathbb{R}^D$, we define the average
\begin{equation*}
\bar{x}=\frac{1}{D}\sum_{i=1}^D x_i
\end{equation*}
and the standard deviation 
\begin{equation*}
\sigma_{\x}=\sqrt{\frac{1}{D}\sum_{i=1}^D (x_i-\bar{x})^2}
 = \frac{1}{\sqrt{D}}\, \Vert \x-{\bf\bar{\x}}\Vert
\end{equation*}
of its components and we set ${\bf\bar{\x}}=(\bar{x},\dots ,\bar{x})$.

We now restrict our attention to $\mathbb{R}^D\backslash Diag$ 
where \[ Diag=\{(x_1,\dots,x_D)\in\mathbb{R}^D|\,x_1=\dots=x_D\}.\]
To $\x\in \mathbb{R}^D\backslash Diag$, we associate the centered and 
reduced variable
\begin{equation*}
{\bf \x^*} = \frac{\x-{\bf\bar{\x}}}{\sigma_{\x}}
           = \sqrt{D}\,\frac{\x-{\bf\bar{\x}}}{\Vert \x-{\bf\bar{\x}}\Vert}
\end{equation*}
Consequently, ${\bf \overline{\x^*}}=0$ and $\sigma_{\bf \x^*}=1$,
and we have
\begin{equation*}
\sigma_{\bf \x^*}^2 = \frac{1}{D} \sum_{i=1}^D (x^*_i)^2 = 1
\Leftrightarrow \, \sum_{i=1}^D(x^*_i)^2 = D 
\end{equation*}
The geometric interpretation of this transform is that ${\bf\x^*}$
lies on the $D$-dimensional hypersphere
$\mathbb{S}^D(\sqrt{D})\subset\mathbb{R}^D$ of radius $\sqrt{D}$
centered at the origin.
 
The correlation between two sample points, $\x=(x_1,\dots ,x_D)$ 
and $\y=(y_1,\dots ,y_D)$, in $\mathbb{R}^D\backslash Diag$ is given by
\begin{equation*}
corr(\x,\y) = 
\frac{\sum_{i=1}^D(x_i-\bar{x})(y_i-\bar{y})}
     {\sqrt{\sum_{i=1}^D(x_i-\bar{x})^2\sum_{i=1}^D(y_i-\bar{y})^2}}\, ,
\end{equation*}
which can also be expressed as
\begin{equation*}
corr(\x,\y) 
= \frac{(\x-{\bf\bar{\x}}) \cdot (\y-{\bf\bar{\y}})}
       {\Vert \x-{\bf\bar{\x}}\Vert \, \Vert \y-{\bf\bar{\y}}\Vert}
= \frac{1}{D} (\x^* \cdot \y^*)
\end{equation*}
where $\x\cdot \y$ stands for the scalar product of $\x$ and $\y$.

\section{A distance based on correlation}
We propose the following correlation-based distance
\begin{equation}
d(\x,\y) = \sqrt{1-(corr(\x,\y))^2}
= \sqrt{1-\frac{({\bf\x^*} \cdot {\bf\y^*})^2}{D^2}}
\end{equation}
for $\x,\y\in \mathbb{R}^D\backslash Diag$. Note that $0\le d(\x,\y)\le 1$.

The following properties of a metric distance
\begin{eqnarray*}
d(\x,\x) &=& 0 \\
d(\x,\y) &=& d(\y,\x) \\
d(\x,\z) &\leq& d(\x,\y) + d(\y,\z),
\end{eqnarray*} 
must be verified.

We have
\begin{equation*}
d(\x,\x) = \sqrt{1-\frac{(\x^* \cdot \x^*)^2}{D^2}}
         = \sqrt{1-\frac{D^2}{D^2}} = 0
\end{equation*}
and, obviously, $d(\x,\y)=d(\y,\x)$. 

The main feature of this distance is that strong correlation
corresponds to small distance. Indeed,
\begin{eqnarray*}
\left [ corr(\x,\y) \right ]^2=1 
&\Leftrightarrow& 
\exists \, \mu\not=0,\delta \in \mathbb{R} \textrm{ s. t. } 
x_i=\mu y_i+\delta, \forall i  \\
&\Leftrightarrow& {\bf\x^*} = \pm{\bf\y^*} \\
&\Leftrightarrow& d(\x,\y)=0.
\end{eqnarray*}
which also means that the distance $d$ is degenerate, since
$d(\x,\y)=0 \not\Rightarrow \x=\y$.

The triangle inequality $d(\x,\z)\leq d(\x,\y)+d(\y,\z)$ requires some
explanations. A preliminary remark is that
\begin{eqnarray*}
d(\x,\y) &=& \sqrt{1-\frac{({\bf\x^*} \cdot {\bf\y^*})^2}{D^2}} 
          =  \sqrt{1-\frac{\left[D\cos(\alpha)\right]^2}{D^2}} 
          =  \sqrt{1-\cos^2(\alpha)} \nonumber \\ 
         &=& \sin(\alpha)
\end{eqnarray*}
where $0\le \alpha\le \pi$ is the angle between ${\bf\x^*}$ and
${\bf\y^*}$.

Replacing ${\bf\y^*}$ by $-{\bf\y^*}$ and ${\bf\z^*}$ by $-{\bf\z^*}$
if necessary, we can assume that the angles $\alpha$ between
${\bf\x^*}$ and ${\bf\y^*}$ and $\beta$ between ${\bf\y^*}$ and
${\bf\z^*}$ belong to $[0,\pi/2]$. Consider the point ${\bf\hat{z}}$
obtained by rotating ${\bf\z^*}$ around the axis defined by
${\bf\y^*}$, into the plane determined by ${\bf\x^*}$ and ${\bf\y^*}$,
but opposite to ${\bf\x^*}$ with respect to ${\bf\y^*}$. The angle
between ${\bf\y^*}$ and ${\bf\hat{z}}$ is still $\beta$. However, the
angle between ${\bf\x^*}$ and ${\bf\hat{z}}$, which equals
$\alpha+\beta$, is greater than the one between ${\bf\x^*}$ and
${\bf\z^*}$. Therefore,
\begin{eqnarray*}
d(\x,\z) 
&\le& \sin(\alpha + \beta) 
  = \sin(\alpha)\underbrace{\cos(\beta)}_{\in [0,1]} +
    \sin(\beta)\underbrace{\cos(\alpha)}_{\in [0,1]}\\
&\le& \sin(\alpha)+\sin(\beta) = d(\x,\y)+d(\y,\z)
\end{eqnarray*}

As previously mentioned, the distance $d$ is degenerate on
$\mathbb{R}^D\backslash Diag$ or on $\mathbb{S}^D(\sqrt{D})$. However,
we obtain a non-degenerate distance on the projective space
$\mathbb{P}^D$ (i.e. the space of lines through the origin in
$\mathbb{R}^D$).

\section{The center of mass}
Onwards, we will assume that all variables are centered and reduced.
Hence, we restrict the sample space to the $D$-dimensional
hypersphere $\mathbb{S}^D(\sqrt{D})\subset \mathbb{R}^D$ of radius $\sqrt{D}$
centered at the origin. We will omit the ${\bf ^*}$ notation.

We compute the center of mass $\g\in \mathbb{S}^D(\sqrt{D})$ of a set of $N$
points $\{\x_j\}_{j=1}^N$ on $\mathbb{S}^D(\sqrt{D})$. By definition, the
center of mass minimizes the average square distance to a set of
points. We therefore want to minimize the expression
\begin{equation}
\label{barycenter}
F(\g) = \frac{1}{N} \sum_{j=1}^N \left[d(\g,\x_j)\right]^2
= 1-\frac{1}{ND^2} \sum_{j=1}^N (\g \cdot \x_j)^2
\end{equation}
under the constraint
\begin{equation}
\label{constraint}
H(\g) = 1 - \frac{1}{D}\, \g \cdot \g = 0
\end{equation}
that $\g$ lies on $\mathbb{S}^D(\sqrt{D})$. 

We solve this problem using the method of Lagrange multipliers.
The gradients of $F$ and $H$ must satisfy 
\begin{equation*}
\nabla F(\g) = \lambda \nabla H(\g)\, ,
\end{equation*}
or equivalently
\begin{equation}
\label{lagrange1}
\frac{\partial}{\partial g_k}F(\g) = 
\lambda \frac{\partial}{\partial g_k} H(\g) \quad (k=1,\dots,D).
\end{equation}
Equation (\ref{lagrange1}) can be rewritten as
\begin{eqnarray}
\frac{1}{ND}\sum_{j=1}^N x_{jk} (\x_j\cdot \g) &=&
\frac{1}{ND}\sum_{j=1}^N \left( x_{jk} \sum_{i=1}^D x_{ji} g_i\right) 
\nonumber \\
&=&\sum_{i=1}^D \left( \frac{1}{ND}\sum_{j=1}^N x_{jk} x_{ji}\right) g_i 
  = \lambda \, g_k
\label{lagrange2}
\end{eqnarray}
If we define the $D\times D$ matrix ${\bf M}=(m_{ik})$ by
\begin{equation*}
m_{ik} =  \frac{1}{ND} \sum_{j=1}^N x_{jk} x_{ji} \quad (i,k=1,\dots,D),
\end{equation*}
then equation (\ref{lagrange2}) becomes
\begin{equation*}
 \sum_{i=1}^D m_{ik} \, g_i = \lambda g_k \quad (k=1,\dots,D) 
\end{equation*}
or equivalently
\begin{equation*}
{\bf M}\g = \lambda \g
\end{equation*}
Thus, minimizing $F$ (eq. \ref{barycenter}) under the constraint $H$
(eq. \ref{constraint}) reduces to finding the eigenvectors of ${\bf
M}$. The eigenvector, correctly normalized in order to satisfy $H$,
for which $F$ is minimum, yields the center of mass of the set of $N$
points $\{\x_j\}_{j=1}^N$ on $\mathbb{S}^D(\sqrt{D})$. The matrix
${\bf M}$ being symmetric, all its eigenvalues are real.
\end{document}